\documentclass[
	aps,
	prx,
	twocolumn,
	amsmath,
	amssymb,
	nofootinbib,
	superscriptaddress,
	floatfix,
	longbibliography
]{revtex4-1}
\usepackage{amsmath}
\usepackage{amssymb}
\usepackage{amsthm}
\usepackage{amsfonts}
\usepackage{listings}
\usepackage{enumerate}
\usepackage{latexsym}
\usepackage{psfrag}
\usepackage{bm}
\usepackage{bbm}
\usepackage[all]{xy}
\usepackage{comment}
\usepackage{hyperref}% add hypertext capabilities
\usepackage{graphicx}
\usepackage{subfigure}
\usepackage{braket}
\usepackage{xcolor}
\usepackage[utf8]{inputenc}
\usepackage[T1]{fontenc}
\usepackage{morefloats}
\usepackage{times}

\def\ie{{\it i.e.},\ }

\newcounter{defcounter}
\begin{document}

\widowpenalty10000
\clubpenalty10000

\newcommand{\cbl}[1]{\color{blue} #1 \color{black}}

\newcommand{\vk}{{\bf k}}

\title{Reciprocal skin effect and its realization in a topolectrical circuit}

\author{
Tobias~Hofmann}
\address{Institute for Theoretical Physics and Astrophysics, University of W\"urzburg, Am Hubland, D-97074 W\"urzburg, Germany}

\author{
Tobias~Helbig}
\address{Institute for Theoretical Physics and Astrophysics, University of W\"urzburg, Am Hubland, D-97074 W\"urzburg, Germany}

\author{
Frank~Schindler}
\address{Department of Physics, University of Zurich, Winterthurerstrasse 190, 8057 Zurich, Switzerland}
\address{Kavli Institute for Theoretical Physics, University of California, Santa Barbara, CA 93106, USA}

\author{
Nora~Salgo}
\address{Department of Physics, University of Zurich, Winterthurerstrasse 190, 8057 Zurich, Switzerland}

\author{Marta~Brzezi\'{n}ska}
\address{Department of Theoretical Physics, Faculty of Fundamental Problems of Technology, Wroc\l{}aw University of Science and Technology, 50-370 Wroc\l{}aw, Poland}
\address{Department of Physics, University of Zurich, Winterthurerstrasse 190, 8057 Zurich, Switzerland}

\author{Martin~Greiter}
\address{Institute for Theoretical Physics and Astrophysics, University of W\"urzburg, Am Hubland, D-97074 W\"urzburg, Germany}

\author{Tobias~Kiessling}
\address{Physikalisches Institut and R\"ontgen Research Center for Complex Material Systems, Universit\"at W\"urzburg, D-97074 W\"urzburg, Germany}

\author{
David~Wolf}
\address{Department of Physics, University of Zurich, Winterthurerstrasse 190, 8057 Zurich, Switzerland}

\author{
Achim~Vollhardt}
\address{Department of Physics, University of Zurich, Winterthurerstrasse 190, 8057 Zurich, Switzerland}

 \author{Anton~Kaba\v{s}i}
\address{Centre of Excellence STIM, University of Split, Polji\v{c}ka cesta 35, HR-21000 Split, Croatia}

\author{Ching~Hua~Lee}
\address{Department of Physics, National University of Singapore, Singapore, 117542.}
\address{Institute of High Performance Computing, A*STAR, Singapore, 138632.}

 \author{Ante~Bilu\v{s}i\'{c}}
\address{Centre of Excellence STIM, University of Split, Polji\v{c}ka cesta 35, HR-21000 Split, Croatia}
\address{University of Split, Faculty of Science, Ru\dj era Bo\v{s}kovi\'{c}a 33, HR-21000 Split, Croatia}

\author{
Ronny~Thomale}
\address{Institute for Theoretical Physics and Astrophysics, University of W\"urzburg, Am Hubland, D-97074 W\"urzburg, Germany}

\author{
Titus~Neupert}
\address{Department of Physics, University of Zurich, Winterthurerstrasse 190, 8057 Zurich, Switzerland}

\begin{abstract}
A system is non-Hermitian when it exchanges energy with its environment and non-reciprocal when it behaves differently upon the interchange of input and response. Within the field of metamaterial research on synthetic topological matter, the skin effect describes the conspiracy of non-Hermiticity and non-reciprocity to yield extensive anomalous localization of all eigenmodes in a (quasi) one-dimensional geometry. Here, we introduce the reciprocal skin effect, which occurs in non-Hermitian but reciprocal systems in two or more dimensions: Eigenmodes with opposite longitudinal momentum exhibit opposite transverse anomalous localization. We experimentally demonstrate the reciprocal skin effect in a passive RLC circuit, suggesting convenient alternative implementations in optical, acoustic, mechanical, and related platforms. Skin mode localization brings forth potential applications in directional and polarization detectors for electromagnetic waves.
\end{abstract}

\date{\today}

\maketitle

%%%%%%%%%%%%%%%%%%%%%%%%%%%%%%%%%%%%%%%%%%%%%%%%%%%%

\begin{figure*}[t]
\begin{center}
\includegraphics[width=0.95\textwidth]{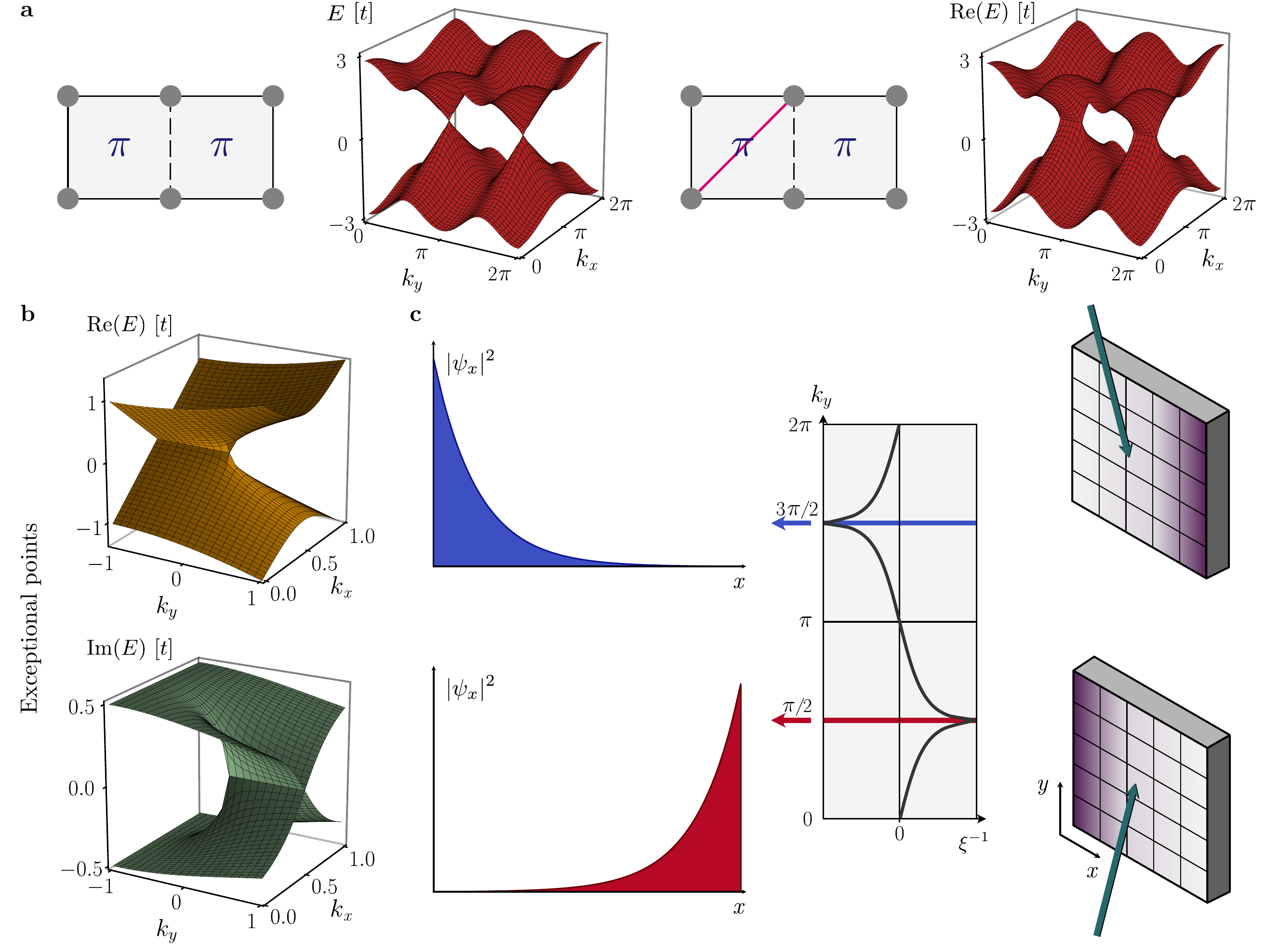}
\caption{
Theory of the reciprocal skin effect.
a) 
Left: unit cell and spectrum of the $\pi$-flux model defined in Eq.~\eqref{eq: pi flux} with periodic boundary conditions. Solid and dashed lines correspond to hopping amplitudes $t$ and $-t$, respectively. Bands touch in two Dirac points.
Right: $\pi$-flux model with a non-Hermitian diagonal hopping (pink) included, as defined in Eq.~\eqref{eq: non-Hermitian H}. Each Dirac point splits into a pair of exceptional points (only the real part of the band eigenvalues is plotted).
b) Real and imaginary part of eigenvalues in the vicinity of an exceptional point, at which two complex eigenvalues are degenerate. 
c) Schematic of the reciprocal skin effect for OBC in $x$ and PBC in $y$ direction: near two opposite momenta, $k_y=\pi/2$ and $k_y=3\pi/2$, \emph{all} eigenstates are exponentially localized with localization length $\xi$ to the left and right of the system, respectively. At $k_y=0,\pi$ the modes are completely delocalized. Right: The reciprocal skin effect could serve as a direction detector for incident electromagnetic waves: dependent on the propagation direction and polarization, a voltage will build up on the left or right edge of the system, as schematically shown in the figure.
}
\label{fig: 1}
\end{center}
\end{figure*}

% +++++++++++++++++++++++++++++++++++++
% Introduction
% +++++++++++++++++++++++++++++++++++++
\section{Introduction}

First developed in the late 1920s
by Felix Bloch and contemporaries~\cite{blochbandtheory}, the
electronic band theory of crystals has
experienced an unprecedented revival of interest and sophistication
through the theoretical and experimental discovery of topological
matter~\cite{klitzing-80prl494,haldane1988model,konig2007quantum}. Whereas
band theory was previously employed in the context of infinite systems, the study of topological
insulators~\cite{RevModPhys.83.1057,RevModPhys.82.3045} and
semimetals~\cite{ArmitageWeylReview18} has reshifted attention to the
interplay of bulk and boundary degrees of freedom, culminating in the
celebrated bulk-boundary
correspondence~\cite{hatsugaiBBC93,prodan2016bulk}. It allows for the prediction of universal boundary phenomena from bulk properties and relies on the fact that changing boundary conditions does not affect the bulk states at large. The predominant
framework for topological crystal matter has been closed systems at zero temperature, whose band spectrum is obtained from a Hermitian Bloch Hamiltonian.

Recently, open systems that do not conserve energy or particle number
have received significant attention. Within band theory, open systems
are modeled by non-Hermitian Hamiltonians. The study of topological
phases supported by such Hamiltonians yields qualitatively novel
phenomena~\cite{Gong18,foatorres19jpm}, such as the occurrence of stable defective degeneracy
points, so-called exceptional points~\cite{lee19prl133903,alvarez18prb121401}. 
In two-dimensional
systems~\cite{fuNonHermitian18}, they are connected by bulk Fermi arcs~\cite{Zhou1009}.
Another
exotic feature resolved through the band theory description of open
systems is the non-Hermitian skin effect~\cite{Shunyu2018prl}: \emph{all} of the eigenstates of a one-dimensional system can localize at one of its
boundaries.
The skin effect constitutes a breakdown of
bulk-boundary correspondence, which is not found in closed
systems~\cite{XiongBBCbreakdown2018,bbbc,PhysRevB.99.201103,BorgniaCitationRequest}. 

The non-Hermitian skin effect has been discussed in
one-dimensional systems, and was initially assumed to necessitate a
non-reciprocal Bloch Hamiltonian usually realized through asymmetric,
direction-dependent hoppings.
Reciprocal versions of the one-dimensional skin effect can be constructed by introducing additional degrees of freedom.
Their protection against hybridization is realized by demanding additional symmetries such as mirror symmetry~\cite{SchnyderPRB2019,okuma2019topological}.
In this work, we present a generalized
skin effect in reciprocal systems, that is enabled by non-Hermiticity
in dimensions higher than one. For simplicity, we constrain ourselves
to two spatial dimensions. The basic idea is that, while a reciprocal
model cannot localize all of its OBC eigenstates at just one
boundary, it may still do so for states of a particular boundary
momentum. Reciprocity then implies that the eigenstates at opposite momenta are localized at opposite boundaries.

The reciprocal skin effect dramatically expands the scenarios in which
anomalous extensive skin mode localization~\cite{PhysRevB.99.201103} can be found in nature. As a
paradigmatic experimental realization within a framework of high
accessibility and tunability, we implement the reciprocal skin effect
within a topolectrical circuit setup. Similar to many other classical
platforms of synthetic topological matter, this roots in the insight
that Berry phase phenomena~\cite{Berry45,PhysRevLett.62.2747} directly
carry over to classical physics~\cite{Haldane:86}, since they do not concern phase space, but parameter space. 
The realization of topological phases in
photonic~\cite{PhysRevLett.100.013904},
mechanical~\cite{lubensky,PhysRevLett.114.114301} and electrical
systems~\cite{ningyuan2015time,PhysRevLett.114.173902,hu2015measurement,LeeTopolectrical18,imhof,PhysRevLett.122.247702,knots,LeeTopolectrical18,interf-nh-ezawa,nh-haldane-ezawa,wang2019topologically} explicates this connection between
topological aspects of quantum and classical systems. The
non-Hermitian band theory of open quantum systems can similarly be
mirrored by descriptions of classical systems that involve gain and
loss. Electric circuits whose circuit Laplacian~\cite{LeeTopolectrical18} is modeled in analogy
to a quantum Bloch Hamiltonian are particularly suited for this: loss
and gain can be directly implemented by resistors and active circuit
elements~\cite{PhysRevLett.122.247702}, respectively, while Hermitian
hybridization elements relate to inductive and capacitive links
between circuit nodes.  We design, describe, and measure~\cite{PhysRevB.99.161114}
a non-Hermitian topolectrical circuit that displays, among exceptional points, the reciprocal skin effect, and
is built entirely from capacitors, inductors, and resistors,
\emph{without} the need for non-reciprocal active elements such as operational amplifiers. Being a
passive circuit network, its convenient translation into alternative
platforms of synthetic topological matter promises ubiquitous
realization and application in
optics, mechanics, and acoustics.

% +++++++++++++++++++++++++++++++++++++
% Theory of the reciprocal skin effect
% +++++++++++++++++++++++++++++++++++++
\section{Theory of the reciprocal skin effect}

To underline the generality of our arguments, we will present the
theory in the language of (non-Hermitian) tight-binding Hamiltonians,
which may represent either a quantum system or the dynamical matrix and response function of a classical system. We start our considerations with the $\pi$-flux tight-binding model on a square lattice. It is characterized by a nearest-neighbor hopping $t$, where exactly one of the four sides of each plaquette has a negative hopping amplitude compared to the three others. These hoppings require a unit cell of two plaquettes of the square lattice [see Fig.~\ref{fig: 1}~a)]. The Bloch Hamiltonian for a system with periodic boundary conditions (PBC) in $x$ and $y$ directions, yielding the momenta $k_x$ and $k_y$, can be written as
\begin{equation}
H_{\mathrm{\pi}}(k_x,k_y)=
t\begin{pmatrix}
2\cos\,k_y&1+e^{-ik_x}\\
1+e^{ik_x}&-2\cos\,k_y
\end{pmatrix}.
\label{eq: pi flux}
\end{equation}
The model has two Dirac-like band touchings at momenta $(k_x,k_y)=(\pi,\pi/2)$ and $(k_x,k_y)=(\pi,3\pi/2)$ [see Fig.~\ref{fig: 1}~a) for the band structure].

We add a non-Hermitian (gain/loss) term as a diagonal hopping through
one of the plaquettes in the unit cell [see Fig.~\ref{fig: 1}~a)]. It
assigns a complex amplitude $i r$ (with $r$ a real number) to
the the process of a particle hopping along the diagonal toward the
upper right, and the exact same amplitude $i r$ to the
reversed process. Hermiticity would require that the latter
process has the complex conjugated amplitude $-i r$. As such, the addition 
\begin{equation}
H(k_x,k_y)=
H_{\mathrm{\pi}}(k_x,k_y)
-ir \begin{pmatrix}
0&e^{ik_y}\\
e^{-ik_y}&0
\end{pmatrix}
\label{eq: non-Hermitian H}
\end{equation}
violates Hermiticity for this diagonal hopping process.
Despite its non-Hermiticitiy, the model still is reciprocal, as it satisfies
$H(k_x,k_y)=H^{\mathsf{T}}(-k_x,-k_y)$ under transposition. 

Hamiltonian~\eqref{eq: non-Hermitian H} has two complex-valued eigenbands with remarkable properties: they  touch (\ie the two eigenvalues have equal real and imaginary part) in two pairs of points. Upon introducing a finite $r$, the Dirac points of the original $\pi$-flux model each split into a pair of these points. Such degeneracy points in non-Hermitian systems are called exceptional points and are the generic band touchings in a space with two parameters (here, $k_x$ and $k_y$). The band structure in the vicinity of an exceptional point is illustrated in Fig.~\ref{fig: 1}~b).

We proceed to consider Hamiltonian~\eqref{eq: non-Hermitian H} with OBC in $x$-direction, which leaves $k_y \in [0,2\pi]$ well defined as a boundary momentum, and define $\tilde{H}(k_y)$ as the Hamiltonian for the strip geometry. While $\tilde{H}^\top(k_y) = \tilde{H}(-k_y)$ guarantees reciprocity of the full system, each instance of $\tilde{H}(k_y)$ \textit{locally} breaks reciprocity for a fixed $k_y$ when seen as a purely one-dimensional model, except for $k_y = - k_y \mod(2\pi)$. The effective 1D model for a given $k_y$ exhibits the non-Hermitian skin effect as discussed in Refs.~\onlinecite{Shunyu2018prl,song2019non}: The eigenstates of a Hermitian system form an orthonormal basis whose squared amplitudes, when summed over all states, are equal on all lattice sites. In a non-Hermitian system this need not be the case, since right (left) eigenstates of a non-Hermitian matrix do not individually form an orthonormal basis. As a result, they can all be localized at only one edge of the system, which defines the skin effect. In our model, the skin effect is realized due to the term proportional to $r$ in Eq.~\eqref{eq: non-Hermitian H}, which renders the hopping probability for going right different from the probability for going left. This leads to an accumulation of \emph{all eigenstates} towards only one edge.

The \emph{reciprocal skin effect} is characterized by a $k_y$ dependence of this localization property. For $k_y \in (0,\pi)$, all OBC bulk modes localize at the right edge, while the localization switches to the left edge for $k_y \in (\pi,2\pi)$. The states are delocalized at $k_y =0, \pi$, because the Hamiltonian $\tilde{H}(k_y)$ is locally reciprocal. We find the strongest localization at $k_y\sim \pi/2, 3\pi/2$, where the initial bulk $\pi$-flux model has its gapless Dirac point. Due to the \textit{global} reciprocity of the Hamiltonian, each localized eigenmode at $k_y$ has a reciprocal partner at $-k_y \mod 2\pi$, which localizes at the opposing edge.
\begin{figure*}[t]
\begin{center}
\includegraphics[width=0.95 \textwidth]{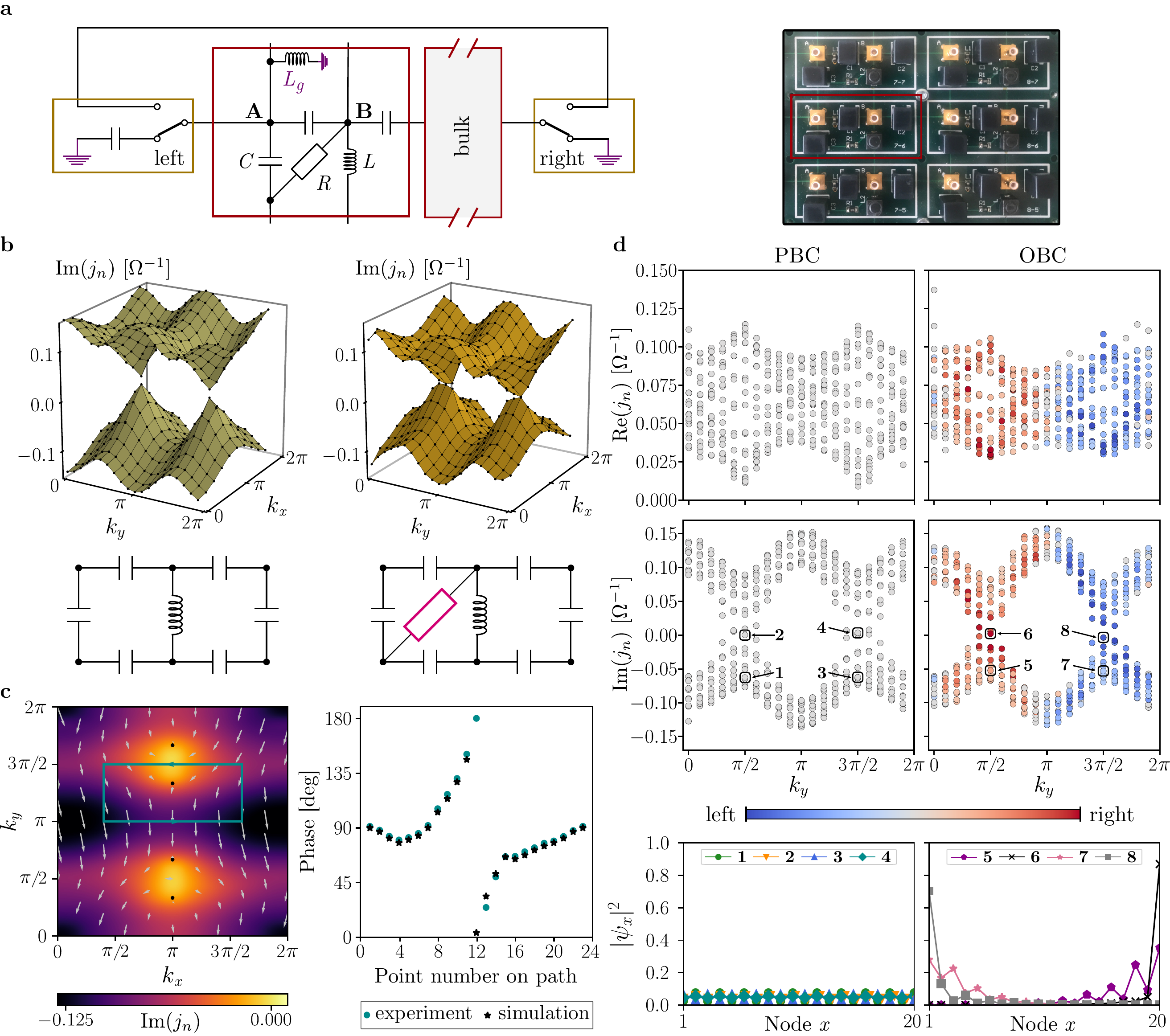}
\caption{
Experimental topolectrical circuit realization of the reciprocal skin effect and exceptional points.
a) Bulk unit cell and boundary terminations of the circuit that realizes the model defined in Eq.~\eqref{eq: non-Hermitian H} and photograph of six unit cells of the assembled circuit board. Appendix C details the connection of this circuit cell to the model in  Eq.~\eqref{eq: non-Hermitian H} using the Laplacian formalism.
b) 
Measured spectra with PBCs of the circuit Laplacian and schematic unit cell of the circuit that corresponds to the  $\pi$-flux model (left) and the non-Hermitian model (right) from Eq.~\eqref{eq: non-Hermitian H} [cf. Fig.~\ref{fig: 1} a)].
Only the imaginary part of the eigenvalues is plotted. The formation of a branch cut from the Dirac point is visible, the ends of which are host to exceptional points.
c)
Left: Imaginary part and phase of the band with smaller imaginary part. The phase of the eigenvalue along the path indicated is plotted in the right panel. It shows a clear jump by $\pi$, indicative of an enclosed exceptional point. (Stars denote a \textsc{LTspice} simulation of the circuit, while dots correspond to the measured spectrum.)
d)
Measured spectra of the circuit Laplacian as a function of $k_y$ for OBC and PBC along the $x$ direction. Localization properties of each eigenstate are indicated by color and determined from the inverse participation ratio (IPR). The bottom panel shows the localization properties of representative individual eigenstates at $k_y = \pi /2$ and $k_y = 3 \pi /2$, both for PBC and OBC.
%Bottom panel show the localization properties of four representative individual eigenstates at $k_y = \pi /2 $.
The fact that \emph{all} OBC and PBC eigenstates differ non-perturbatively constitutes the reciprocal skin effect, with those  around $k_y\sim\pi/2$ right-localized and those around $k_y\sim3\pi/2$ left-localized.
}
\label{fig: 2}
\end{center}
\end{figure*}

The reciprocal skin effect in two dimensions is fundamentally different to a doubled and reciprocity-enhanced 1D skin effect, as realized for instance by two reciprocity-reversed Hatano-Nelson chains~\cite{hatano-nelson}. To conceptualize this, consider an analogy to the 2D Chern insulator and the 3D Weyl semimetal~\cite{Shuichi_Murakami_2007}.
Coupling two time-reversed copies of a Chern insulator with chiral edge states constitutes the $\mathbb{Z}_2$ topological insulator with helical edge states and restores time-reversal symmetry~\cite{PhysRevLett.95.146802}. In a similar fashion, one can create an overall reciprocal system out of two Hatano-Nelson chains, whose exponentially localized modes are not protected by symmetry.
Only by the demand of additional symmetries which square to $-1$, the precise notion of a $\mathbb{Z}_2$ skin effect is defined~\cite{okuma2019topological}.
In distinction to this symmetry enhancement, we consider dimensional enhancement. 
The Weyl semimetal in 3D can be composed of ``slices'' of momentum space, which are characterized by a 2D Chern insulator. 
By analogy, we extend the 1D (non-reciprocal) skin effect to two dimensions, arriving at the reciprocal skin effect, with momentum space slices that host the 1D skin effect.
Slices at opposite momenta are connected by the reciprocity transformation. In contrast to one-dimensional systems, hybridization of those reciprocal partners is intrinsically prevented by the translational symmetry of the two-dimensional system.

% +++++++++++++++++++++++++++++++++++++
% Topolectrical circuit realization
% +++++++++++++++++++++++++++++++++++++
\section{Experimental topolectrical circuit realization}

In principle, various non-Hermitian, \ie lossy, classical systems
could be deliberately tailored to study the physical effects outlined
above. Topolectrical circuits in particular, however, offer an
important advantage due to the only mild limitation imposed by local
connectivity contstraints otherwise typical to many other metamaterial
settings. Most importantly, a connection between the leftmost and rightmost site of a circuit can simply be toggled on/off to change between PBC and OBC. This way, the breakdown of bulk-boundary correspondence can be studied most directly.  

We experimentally implemented a circuit which realizes the non-Hermitian but reciprocal model~\eqref{eq: non-Hermitian H} through its response function within linear circuit theory, the circuit Laplacian matrix $J (\omega)$, which takes the role of the reciprocal Hamiltonian previously introduced. It connects the input currents $I_a (\omega)$ at node $a$ of the system to the voltages $V_b (\omega)$ measured at $b$ via Kirchhoff's law, represented in the frequency domain by $I_a (\omega)=J_{ab}(\omega) V_b(\omega)$. Here we consider a fixed excitation frequency $\omega$, the summation over the repeated index $b$ is implied. (See Supplemental Information for the theory on circuit Laplacians.)

In a Hermitian circuit, all eigenvalues of $J(\omega)$ are purely imaginary [$iJ(\omega)$ is a Hermitian matrix], while the inclusion of resistors generates complex eigenvalues.
Interpreting $J (\omega)$ at a fixed frequency $\omega_0$ as a hopping matrix, one observes that the sign change of $t$ necessary to implement the $\pi$-flux model is achieved by connecting one of the three bonds surrounding a plaquette in a square lattice with an inductor and three with a capacitor [see Fig.~\ref{fig: 2}~a) and~b)]. We fabricated a circuit with $10 \times 20$ unit cells using this model. (See Supplemental Information for the exact specifications of the circuit.) 

Assuming translation invariance (realized to the accuracy of the circuit element specifications), we can represent the voltages and input currents in terms of Fourier modes in reciprocal space. This leads to the $k$-space representation of the Laplacian as well as its voltage eigenmodes and allows for the definition of a complex-valued admittance band structure. The latter can be understood as a complex mapping from wave vector $k$ to admittance eigenvalues of the Laplacian. In a measurement, we can decompose the voltage response to an external current excitation and find the eigensystem of the Laplacian for both periodic and open boundary conditions. Figure~\ref{fig: 2}~b) (left) shows the measured bands including the two Dirac-cone band touchings expected for the $\pi$-flux model.

Unlike in experimental realizations of the non-reciprocal skin
effect~\cite{bb-breakdown,nhssh-metamat,qwalks1}, the reciprocal skin
effect can be effected without active elements. For that, we added
resistive couplings across the diagonal of every other row of
plaquettes, which is analogous to the non-Hermitian term proportional
to $r$ in Eq.~\eqref{eq: non-Hermitian H}. Figure~\ref{fig: 2}~b) (right) shows the measured bands.
The Dirac points are broadened into branch cuts, each
of which spans between a pair of exceptional points. To demonstrate this, we plot the phase of the eigenvalue for the band $j_n(\omega_0,k_x,k_y)$ with smaller imaginary part along a closed path in Fig.~\ref{fig: 2}~c). The observed winding and phase jump by $\pi$ is direct evidence that the path encircles an exceptional point in momentum space, which is topologically stable exactly through this half-integer winding number of the band eigenvalue around it.

Having confirmed that the circuit realizes the desired physics of an
exceptional point band structure with PBC, we now present measurements
with OBC in $x$-direction to demonstrate the reciprocal skin
effect. Edge terminations are chosen such that the circuit grounding
does not introduce undesired off-sets due to a change in the total
node conductance at the edge sites, see
Fig.~\ref{fig: 2}~a). The measured $k_y$-resolved spectra are shown in Fig.~\ref{fig: 2}~d), in comparison to an
equivalent representation of the data for PBC. The eigenvalues with
small imaginary part around $k_y\sim \pi/2, 3\pi/2$ indeed show the expected
reorganization from a spectrum with two exceptional points in the bulk
towards much fewer states with OBC -- a breakdown of the Hermitian
bulk-boundary correspondence. The reciprocal skin
effect is encoded in the coloring of the data points: red and blue
dots correspond to right and left localized eigenstates. (Note that
through the measurement of the full matrix $J(\omega_0)$ not only
do we have access to its spectrum, but also to all of its eigenstates.) A degree of localization of eigenstates is quantified by the inverse participation ratio (IPR)~\cite{ipr}. Remarkably, \emph{all} states near $k_y\sim \pi/2$ are right-localized, while \emph{all} states near $k_y\sim 3\pi/2$ are left-localized.

% +++++++++++++++++++++++++++++++++++++
% Discussion
% +++++++++++++++++++++++++++++++++++++
\section{Discussion}

We have introduced and experimentally demonstrated the concept of a
reciprocal skin effect, where the breakdown of bulk-boundary
correspondence occurs in the absence of any non-reciprocal
coupling. Instead of having extensive mode accumulation all along one
boundary, an equal number of eigenmodes localizes along opposite
boundaries, with the direction of localization tied to the momentum
component parallel to the boundary. Key to their realization is the
gain/loss associated with couplings across different sites, which
effectively behave like non-reciprocal couplings at a fixed transverse
momentum. The reciprocal skin effect can in principle exist in two or higher dimensions when the non-Hermitian reciprocal couplings connect different internal degrees of freedom and the momentum space structure protects the skin modes from hybridization.

We observed the reciprocal skin effect in an electric circuit with
solely passive linear circuit elements. Key to its realization are
serial resistors, which are natural candidates for inter-site
non-Hermitian loss, instead of active non-reciprocal elements such as
operation amplifiers necessary for the non-reciprocal conventional
skin effect. The breakdown of bulk-boundary correspondence becomes
evident by comparing the PBC system with exceptional points with the markedly different OBC case, featuring oppositely localized skin modes. Our circuit, and more generally the reciprocal skin effect, facilitates novel functionalities when coupled to electromagnetic waves. For instance, it lends itself to potential applications for polarization and direction detectors for electromagnetic waves, where differently directed or polarized input signals are substantially accumulated towards opposite directions.
%%%%%%%%%%%%%%%%%%%%%%%%%%%%%%%%%%%%%%%%%%%%%%%%%%%%

\begin{acknowledgments}
We thank C. Coulais, S. Huber, V. Vitelli, and Z. Wang for fruitful
discussions, as well as M. Hengsberger for guidance with the
measurement setup and F. Natterer for lending measurement
equipment.  The circuit simulations have been performed by the use of \textsc{LTspice}. The work in W\"urzburg is funded by the Deutsche
Forschungsgemeinschaft (DFG, German Research Foundation) through
Project-ID 258499086 - SFB 1170 and through the W\"urzburg-Dresden
Cluster of Excellence on Complexity and Topology in Quantum Matter --
\textit{ct.qmat} Project-ID 39085490 - EXC 2147.
FS was supported by the Swiss National Science Foundation (grant number: 200021\_169061), the National Science Foundation under Grant No. NSF PHY-1748958, and by the Heising-Simons Foundation. MB was supported by the Polish National Agency for Academic Exchange. We further acknowledge support by ERC-StG-Neupert-757867-PARATOP.
This research was partially supported under the project STIM – REI, Contract Number: KK.01.1.1.01.0003, a project funded by the European Union through the European Regional Development Fund – the Operational Programme Competitiveness and Cohesion 2014-2020 (KK.01.1.1.01).
\end{acknowledgments}

%%%%%%%%%%%%%%%%%%%%%%%%%%%%%%%%%%%%%%%%%%%%%%
% APPENDIX
%%%%%%%%%%%%%%%%%%%%%%%%%%%%%%%%%%%%%%%%%%%%%%

\appendix

\section{Necessary conditions for the reciprocal skin effect}
The reciprocal skin effect can occur in any physical system in which reciprocity can be understood as 
a perfect cancellation of reciprocity-breaking of two oppositely non-reciprocal subsystems. As such, when separated into these subsystems, e.g., specific momentum sectors, a reciprocal system may appear effectively non-reciprocal. While this general observation pertains to most reciprocal systems, we restrict our study to non-Hermitian systems and their individual momentum subsectors, since they allow for the definition of topological invariants as key ingredients of bulk-boundary correspondence.

To illustrate the idea, we adopt the language of  Hamiltonian matrix elements acting on localized degrees of freedom (nodes).
On a finite lattice of at least two dimensions, a coupling $\lambda$ between two nodes $a,b$ displaced by a distance $Y$ transverse to the boundary gives rise to hopping elements $H_{ab}=\lambda \, e^{ik_yY}$ and $H_{ba}=\lambda \, e^{-ik_yY}$ in the Hamiltonian matrix by reciprocity defined as $H(k_y)=H^{\top}(-k_y)$. If $\lambda$ were real, $H_{ab}=H^*_{ba}$ and the Hamiltonian would not pick up any non-Hermiticity, that is necessary for the skin effect to arise. Suppose $\lambda$ is complex, \ie $\lambda=\lambda_0 \, e^{i\theta}$, where the complex phase $\theta$ corresponds to a phase difference of impedance connections in the circuit interpretation. This type of coupling gives rise to a contribution
\begin{equation}
\lambda_0 \, e^{i\theta}\,\left[\cos(k_yY)\,\sigma_x+\sin(k_yY)\,\sigma_y\right]
\end{equation}
to the Hamiltonian in the $(a,b)$ basis. For $\theta \neq m  \pi$, $m\in \mathbb{Z}$, it is manifestly non-Hermitian. By regarding $k_y$ as a parameter and not a variable of the theory, we induce non-reciprocity on individual slices of $k_y$ in the term proportional to $\sigma_y$. In particular, the effective non-reciprocity occurs with a nonzero coupling distance $Y$ and its complex-valued prefactor $\lambda$. Due to the oddness of $\sin(k_yY)$ in front of $\sigma_y$, states with opposite $k_y$ experience reversed effective non-reciprocity. As a result, these eigenstates at fixed opposite $k_y$ accumulate along opposite boundaries as a consequence of the skin effect.

In common reciprocal system setups such as RLC circuits, this reciprocal skin effect is usually not observed because excitation signals are designed to be almost invariably reciprocal, consisting of an equal superposition of $+ k_y$ and $-k_y \mod 2\pi$ modes. Skin mode accumulation in such systems can however be physically observed, if the input signals are polarized to have unequal $\pm k_y \mod 2\pi$ weights. This is detailed in methods Sec.~\ref{sec: responsesupplementary}, where differently directed signals, created by suitably designed phase differences between sublattices, are demonstrated to induce voltage responses with a non-reciprocal directional preference.

\section{Non-Hermitian circuit theory}
In a passive circuit network of capacitors (C), inductors (L), and resistors (R), currents and voltages are linearly related through a discretized version of the Laplacian operator as a second order spatial derivative. However, energy in such a circuit system is not necessarily conserved, as it can dissipate in an irreversible heating process occuring in resistors. In contrast to purely capacitive or inductive couplings, which oppose the \textit{change} of electric currents or voltages, a resistor opposes the \textit{flow} of an electric current and is, as such, intrinsically non-Hermitian. To be able to analyze the present circuit setup, which involves all of the passive components R, L, C, we rely on non-Hermitian linear circuit theory.

Define $V_a$ and $I_a$ to be the voltage and external input current on node $a$ of a circuit network. Using Kirchhoff's and Ohm's laws, we obtain a coupled system of differential equations for the circuit,
\begin{equation}
\dot{I}_a=\Gamma_{ab} \, \ddot{V}_b+\Sigma_{ab} \, \dot{V}_b+\Lambda_{ab} \, V_b,
\label{dotI}
\end{equation}
where $\Gamma_{ab},\Sigma_{ab}$ and $\Lambda_{ab}$ are the reduced Laplacian matrices of capacitances, conductances and inverse inductances, and the summation over repeated indices is implied. The diagonal components $a=b$ of the Laplacians are defined by 
\begin{equation}
X_{aa}=-X_{a0}-\sum_{b=1,2,\cdots}X_{ab},\qquad X\in\{\Gamma,\Sigma,\Lambda\},
\end{equation}
including the circuit elements $X_{a0}$ between node $a$ and the ground. 

A Fourier transformation of Eq.~\eqref{dotI} from the time to frequency domain results in
\begin{equation}
I_a=\left(i\omega \, \Gamma_{ab}+\Sigma_{ab}-\frac{i}{\omega} \, \Lambda_{ab}\right)V_b =J_{ab}(\omega) \, V_b ,
\label{I2}
\end{equation}
where we defined $J_{ab}(\omega)$ as the (grounded) circuit Laplacian~\cite{LeeTopolectrical18}. Note that $\omega$ is treated as a parameter of the system which is fixed by the external AC driving frequency. 

A natural observable in a circuit is the impedance response $Z_{a 0}(\omega)$, which is the ratio of the voltage at node $a$ measured with respect to ground due to an input current $I_{j}=I_0\,\delta_{j,a}$ that enters through $a$ and exits through ground. Mathematically, $Z_{a0}(\omega)$ simply involves the inversion of Eq.~\eqref{I2}
\begin{align}
Z_{a0}(\omega)&= \frac{V_a}{I_0} 
= \sum_j\frac{G_{aj}(\omega) \, I_i}{I_0}
= G_{aa}(\omega) \nonumber \\
&= \sum_n\frac{\psi_{n,a} \, \phi_{n,a}^{*}}{j_n(\omega)},
\label{impedance1}
\end{align}
where $J_{ab}(\omega)=\sum_n j_n(\omega) \psi_{n,a} 
\phi^*_{n,b}$ defines the spectral representation of the Laplacian with its right and left eigenvectors, $\psi_n$ and $\phi_n$. The frequency dependence of the Laplacian eigenvectors remains implicit. 
As the inverse of the Laplacian, the Green's function $G_{ab}(\omega)=\sum_n j^{-1}_n(\omega) \psi_{n,a} \, \phi_{n,b}^{*}$ contains the voltage response to an external current excitation and fundamentally determines both the excitation pattern of individual eigenmodes and the circuit's impedance profile with frequency. The grounding impedances are given by the diagonal elements of $G$.

The off-diagonal components of the Green's function are accessible using an arrangement of measurements similar to that of the impedance response $Z_{a 0}(\omega)$. We feed a current $I_a$ at node $a$ and measure the voltage response $V_b^{(a)}$ at all the other nodes. By repeating this for all input nodes, the Green's function can be reconstructed as
\begin{align}\label{eq:Greens Function Measurement}
G_{a b} = \frac{V_b^{(a)}}{I_a} .
\end{align}
In the circuit formalism, the Greens function as a direct observable contains full information on admittance eigenvalues and eigenmodes of the Laplacian, which can be extracted through numerical diagonalization. The procedure to measure the Green's function can be simplified in periodic models using spatial Fourier transform~\cite{PhysRevB.99.161114} (see Sec.~\ref{appx:Experimental Implementation}).

\section{Theoretical circuit analysis}
\subsection{Derivation of the circuit Laplacian}
We realize the reciprocal skin effect in an electrical circuit whose unit cell is shown in Fig.~2~a), with the conceptually important elements of each plaquette shown in Fig.~2~b).  The two-point Laplacian of a capacitor is given by
\begin{align}
\begin{pmatrix}
I_{\text{in},1}\\I_{\text{in},2} 
\end{pmatrix}
= i\omega \left[ C 
\begin{pmatrix}
1 & -1 \\ -1 & 1
\end{pmatrix}
\right]
\begin{pmatrix}
V_1 \\ V_2
\end{pmatrix} .
\end{align}
Under omission of the prefactor $i \omega$, the Laplacian matrix is symmetric and Hermitian. The Laplacian for an inductor takes a similar form
\begin{align}
\begin{pmatrix}
I_{\text{in},1}\\I_{\text{in},2} 
\end{pmatrix}
= i\omega \left[ - \frac{1}{\omega^2 L}
\begin{pmatrix}
1 & -1 \\ -1 & 1
\end{pmatrix}
\right]
\begin{pmatrix}
V_1 \\ V_2
\end{pmatrix} ,
\end{align}
that differs in the frequency-dependent admittance prefactor and, importantly, in the overall sign. At the resonance frequency $\omega_0 = 1/\sqrt{L C}$ of an $LC$ resonator, the inductor effectively acts a negative capacitor, where $-1/(\omega_0^2 L) = -C$ in front of the Laplacian matrix. Therefore, one circuit plaquette of the $\pi$-flux model consists of three capacitors (representing $t$) and an inductor (representing $-t$ at resonance), all of which are Hermitian and reciprocal. At nodes of sublattice $A$, the diagonal contributions in the Laplacian add to $4 C$, whereas for $\omega = \omega_0$ the capacitive and inductive contributions cancel each other in the diagonal term at sublattice $B$. To avoid the sublattice asymmetry contributing to a $\sigma_z$ term, we ground nodes of type $A$ with an inductor $L_g \approx L/4$ such that the diagonal contribution to the Laplacian vanishes for both sublattices at resonance frequency.
In total, the circuit Laplacian in momentum space is given by
\begin{align}
J_\pi(k_x,k_y&) = i\omega \bigg[
	-\begin{pmatrix}
	2 C \cos(k_y)  & C(1 + e^{-i k_x}) \\
	C(1+e^{i k_x}) & 2/(\omega^2 L) \cos(k_y) 
	\end{pmatrix} \notag \\
	&+
	\begin{pmatrix}	
	4 C - 1/(\omega^2 L_g) & 0 \\
	0 & 2 C - 2/(\omega^2 L)
	\end{pmatrix}
\bigg]
\end{align}
resembling the $\pi$-flux tight-binding model for $\omega \rightarrow \omega_0$, where the second term vanishes.

Additional to $J_\pi(k_x,k_y)$, we introduce a resistor connecting nodes $A$ and $B$ of adjacent unit cells in $y$-direction [see Fig.~2~a),~b)]. Its admittance representation reads
\begin{align}
\begin{pmatrix}
I_{\text{in},1}\\I_{\text{in},2} 
\end{pmatrix}
&= i\omega \left[ - \frac{i}{\omega R}
\begin{pmatrix}
1 & -e^{i k_y} \\ -e^{-i k_y} & 1
\end{pmatrix}
\right]
\begin{pmatrix}
V_1 \\ V_2
\end{pmatrix} \nonumber \phantom{(12)}\\
&\equiv J_{r} \begin{pmatrix}
V_1 \\ V_2
\end{pmatrix}.
\end{align}
The total Laplacian is given by $J = J_{\pi} +J_r$. With $i \omega$ factored out, the added Laplacian $J_r$ is non-Hermitian and breaks time-reversal symmetry. For an arbitrary circuit network described by $J(k_y,k_y)$, reciprocity is defined as $J^\top(k_x, k_y) = J(-k_x, -k_y)$. Thus, a resistor is a reciprocal circuit element. However, if one considers a fixed $k_y$ slice of the model, reciprocity is broken as $e^{i k_y} \neq e^{-i k_y}$ for arbitrary $k_y\neq 0,\pi$. Consider OBC in $x$-direction, such that the $y$-direction preserves translational invariance and $k_y$ remains to be well-defined. We find the non-Hermitian skin effect with an extensive number of boundary localized modes, for a range of momenta $k_y$. The emergence of the skin effect on the level of the tight-binding model is analyzed in the following.

\subsection{Hermitian part of the model}
We omit the unit matrix contribution to the circuit Laplacian $J$ and interchangeably consider the tight-binding model that it induces. The Hermitian part of the model is based on the $\pi$-flux model with the Hamiltonian
\begin{align}\label{eq:pi flux Ham}
H_\pi = \left[1+\cos(k_x)\right]\sigma_x + \sin(k_x) \, \sigma_y + 2 \cos(k_y) \, \sigma_z.
\end{align}
It features two Dirac band touchings, which are located on the boundary of the Brillouin zone at $(k_x, k_y) = (\pi, \pi/2), (\pi,3\pi/2)$. 
Applying open boundary conditions to the $\pi$-flux model in $y$-direction in a cylindrical geometry results in a pair of counter-propagating modes in the boundary Brillouin zone, as the two Dirac cones are projected onto the same point. There are no topological boundary modes in this configuration. Open boundaries in $x$-direction preserve the separation of the Dirac crossings at $k_y = \pi/2, 3\pi/2$. If we treat $k_y$ as a parameter instead of as a variable in Eq.~\eqref{eq:pi flux Ham}, we end up with an effectively one-dimensional hopping model, that is form-invariant to a hopping model on a 1D chain with an additional chiral symmetry breaking mass of $ m = 2 \cos(k_y)$ parametrized by $k_y$. As the amplitudes of the intracell and intercell hoppings of the chain are identical, there will be no topological edge states for open boundary conditions and in particular no flat band joining the two projected Dirac cones in the $k_y$ boundary Brillouin zone.

\subsection{Full non-Hermitian model}
We now introduce a non-Hermitian extension to the $\pi$-flux model by adding a diagonal coupling across one of the two plaquettes in the unit cell and arrive at the total Hamiltonian
\begin{align}
H= H_\pi - i r\,\cos(k_y)\, \sigma_x + i r\,\sin(k_y)\, \sigma_y,
\end{align}
which is related to the circuit Laplacian by $H = [J(\omega_0)-i r \mathbbm{1}]/(i\omega_0 C)$ and $r=1/(\omega_0 R C) $. It is reciprocal, $H(k_x, k_y)^\top = H(-k_x,-k_y)$, in agreement with the fact that it was constructed out of passive elements only. We can tune the strength of the non-Hermiticity by the resistance $R$, which is chosen such that $r = 1$ in our implementation.
The eigenvalues of the Hamiltonian $H$ are complex-valued and the Dirac cones each split into two exceptional points, which are located at $k_x=\pi$ and $k_y = 1/2 \, \arccos(r^2/2-1)$. The exceptional points are band closings of a complex-valued band structure where the corresponding Hamiltonian is defective, \ie the two eigenvectors coalesce to one and the matrix is a non-diagonalizable Jordan block. The band closing point for $k_x$ remains unchanged upon the added non-Hermiticity. For $r=1$, the four PBC exceptional points are pinned to $k_y = \pm \pi/3, \pm 2 \pi/3 \mod 2\pi$. 

\begin{figure}[t]
\begin{center}
\includegraphics[width=0.95\linewidth]{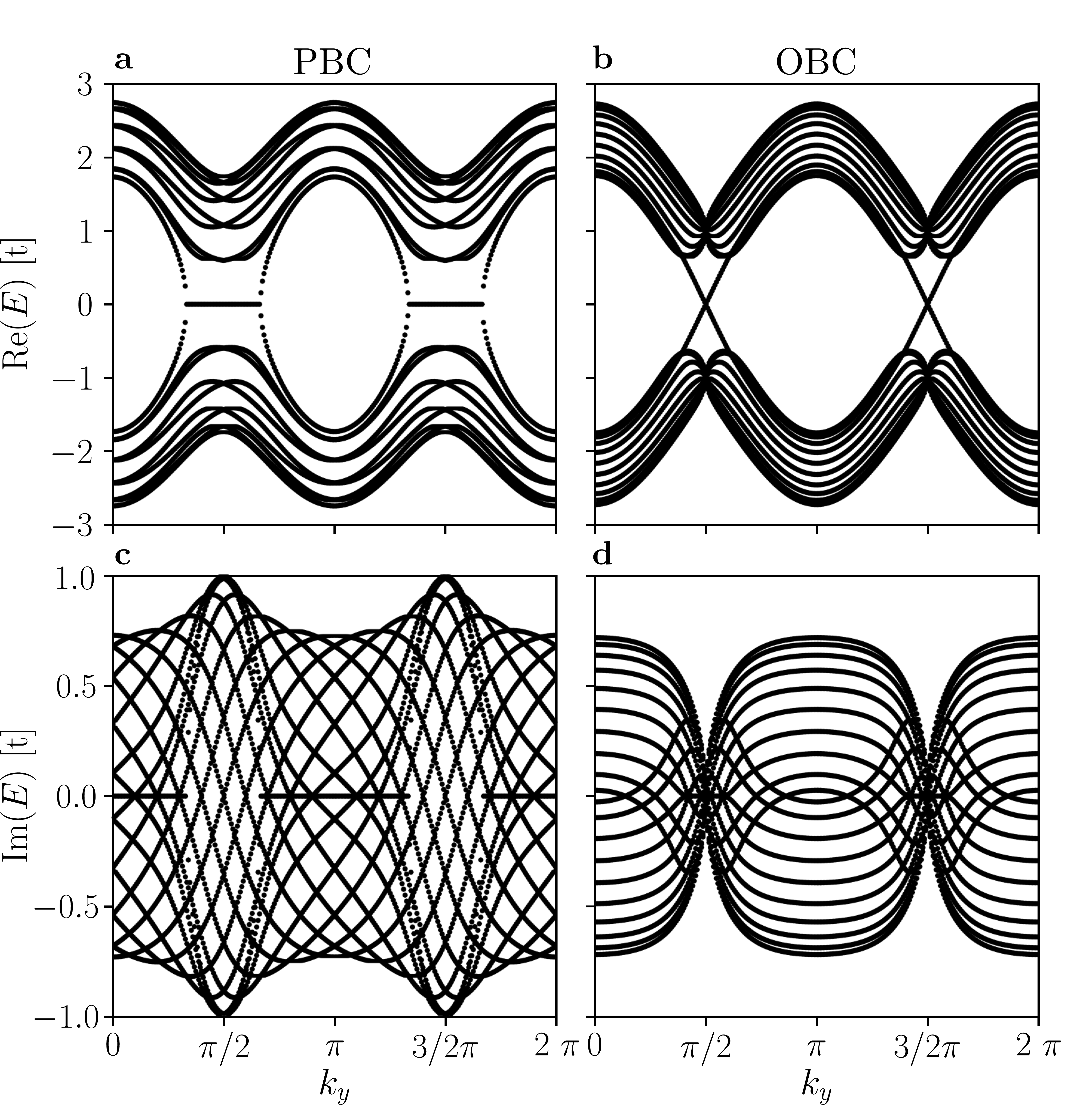}
\caption{
Numerical calculation of the complex eigenspectrum of the non-Hermitian, non-reciprocal Hamiltonian in eq.~\eqref{eq:full_non_hermitian_hamiltonian} for $r = 1$. \textbf{(a)} Real part of the PBC spectrum, \textbf{(b)} real part of the OBC spectrum, \textbf{(c)} imaginary part of the PBC spectrum, \textbf{(d)} imaginary part of the OBC spectrum. There is a non-trivial spectral flow from PBC to OBC, which manifests the breakdown of bulk-boundary correspondence associated with the extensive localization of eigenmodes.
}
\label{fig:num_spectrum}
\end{center}
\end{figure}

The effectively one-dimensional model obtained by treating $k_y$ as a parameter is extended by two non-Hermitian terms resulting in 
\begin{align}\label{eq:full_non_hermitian_hamiltonian}
H_{k_y}(k_x) = &\, \big[1+\cos(k_x) - i r_x \big]\,\sigma_x  \nonumber \\
&\,+ \big[\sin(k_x) + i r_y\big]\,\sigma_y + m \, \sigma_z
\end{align}
with $r_x = r \cos(k_y)$, $r_y = r \sin(k_y)$ and $m = 2 \cos(k_y)$.
The spectrum of $H_{k_y}(k_x)$ for periodic and open boundary conditions is shown in Fig.~\ref{fig:num_spectrum} for $r = 1$. The spectra for OBC and PBC are manifestly different demonstrating the non-trivial spectral flow.

If we fix $k_y$ to one particular value, which means taking $r_x$, $r_y$ and $m$ as constants, $H_{k_y}(k_x)$ breaks reciprocity resulting from the term $i r_y \sigma_y$. The combined breaking of reciprocity and Hermiticity in this model gives rise to the one-dimensional skin effect for $r_y\neq 0$~\cite{Shunyu2018prl,XiongBBCbreakdown2018,PhysRevB.99.201103}.

The bulk modes of the OBC Hamiltonian are localized on one boundary of the system with $\psi_x \sim e^{-x/\xi}$, where 
\begin{align}
\xi^{-1} &= \frac{1}{4} \ln\left(\frac{(1+r_y)^2+r_x^2}{(1-r_y)^2+r_x^2}\right) \nonumber \\
&= \frac{1}{4} \ln\left(\frac{1+r^2+2 r \sin(k_y)}{1+r^2-2 r \sin(k_y)}\right).
\end{align}
The localization length $\xi$ is positive, if $k_y \in \, (0,\pi)$ and negative if $k_y \in \, (\pi,2\pi)$ in the boundary Brillouin zone with open boundaries along $x$. This leads to left edge localized modes, if $k_y \in \, (0,\pi)$ and to right edge localization, if $k_y \in \, (\pi,2\pi)$. The strongest localization is found at the original positions of the Dirac crossings in the boundary Brillouin zone,  $k_y = \pi/2, 3\pi/2$.  At those points, $\xi$ vanishes for $r\rightarrow 1$ leading to infinite localization, which is accompanied by exceptional points in the OBC spectrum of the model. 
The localization inherits a reciprocal symmetry from the full model as $\xi(-k_y) = -\xi(k_y)$. A state at $k_y$ has a reciprocal partner mode at $-k_y \mod 2\pi$, which possesses the same absolute localization length, but is localized on the opposite edge. Combining those partners preserves reciprocity in the full model. For $k_y = 0$ or $\pi$, the localization length diverges. The corresponding bulk eigenstates are delocalized and bulk-boundary correspondence is restored. This roots in the fact that $r_y = 0$ at $k_y = 0$ or $\pi$ and $H_{0,\pi}(k_x)$ is reciprocal leading to a matching of the PBC and OBC spectrum.

Due to the inclusion of non-Hermitian terms, topological states can emerge in the full Hamiltonian, that have not been present in the initial Hermitian $\pi$-flux model. Our discussion closely follows Ref.~\onlinecite{Shunyu2018prl}. To study the topological aspects in the system with $N$ unit cells, we perform a \textit{non-unitary} transformation $H_{k_y}' = S^{-1}H_{k_y}S$ with $S= \text{diag}(1, a,a,a^2,a^2, \dots,a^{N-1}, a^N)$ and $a= \sqrt{(1-i r_x -r_y)/(1-i r_x +r_y)}$, where open boundary conditions are implied. The transformed Hamiltonian under application of PBCs can be rewritten as a reciprocal Su-Schrieffer-Heeger (SSH) chain with complex coefficients,
\begin{align}
H_{k_y}' =  \big[t_0+\cos(k_x) \big]\,\sigma_x + \big[\sin(k_x) \big]\,\sigma_y + m \, \sigma_z,
\end{align}
where $t_0 = \sqrt{(1-i r_x)^2-r_y^2} \in \mathbb{C}$. SSH-type edge modes exist on both edges of the system, if $|t_0|<1$, which translates to a topological regime for
\begin{align}
\cos(k_y) < \sqrt{\frac{1}{2} - \left(\frac{r}{2}\right)^2},
\end{align}
parametrized by the momentum $k_y$. The bulk spectrum of $H_{k_y}'$ is gapped, if either $t_0 \neq 1$ or $m\neq 0$. The gap closes at $r=\sqrt{2}$ for $k_y = \pi/2$, which marks the topological phase transition in $r$. If $r\geq\sqrt{2}$, there are no topological states. As the spectra of $H_{k_y}'$ and $H_{k_y}$ are identical, the total Hamiltonian with OBCs in $x$-direction exhibits a bulk gap for all $k_y$ in the boundary Brillouin zone for $r\neq \sqrt{2}$. The topological phase transition does in general not occur at the PBC band closings, but is instead parametrically displaced due to the non-Hermiticity. Only for the special choice of $r=1$, the two conditions coincide, resulting in topological transitions at the PBC exceptional points for $k_y = \pm \pi/3, \pm 2 \pi /3 \mod 2\pi$. The topological boundary modes disperse with $\pm m = \pm 2 \, \cos(k_y)$ and assume a linear dispersion at the original location of the Dirac crossings in the non-Hermitian model at $k_y = \pi/2, 3\pi/2$ in the boundary Brillouin zone. Note that the bulk is gapped at those points, while the edge states cross at zero energy. The analysis shows that SSH-type modes can coexist with skin modes in the non-Hermitian model with open boundaries along $x$.

\section{Experimental circuit implementation}
\label{appx:Experimental Implementation}
The circuit board was fitted with metallized polypropylene film capacitors (Vishay MKP1837410161G, 100nF 160V) implementing $C$ and SMD power inductors of 33$\mu$H (Bourns SRR0604-330KL, max. 250m$\Omega$ DC resistance) for $L$ and 8.2$\mu$H (TDK MLF2012E8R2JT000, max. 700 m$\Omega$ DC resistance) for $L_\text{g}$. For $R$ we use the $18\,\Omega$ resistor (Yageo RC1206FR-0718RL). The capacitors and resistors have a tolerance of 1\% and were therefore not characterized. All inductances were pre-characterized with the Hameg HM8118 LCR Brige to obtain tolerances of below 1\% of the nominal component values. 

The measurement was conducted at the frequency of $\text{87.25}\,\text{kHz}$, which was identified as the resonance frequency $f_0 = \omega_0/(2\pi)$. This frequency differs from the nominal value of $1/\sqrt{LC}$ as parasitic effects in the inductors and capacitors shift the resonance. To find the measurement frequency, a frequency sweep of the voltage at a node $A$ and a node $B$ due to a current excitation was done and compared to an \textsc{LTspice} simulation of the circuit including parasitic resistances. On the basis of features such as local minima and maxima in the frequency sweep the resonance frequency could be identified.

\begin{figure*}[t]
\begin{center}
\includegraphics[width=0.95\textwidth]{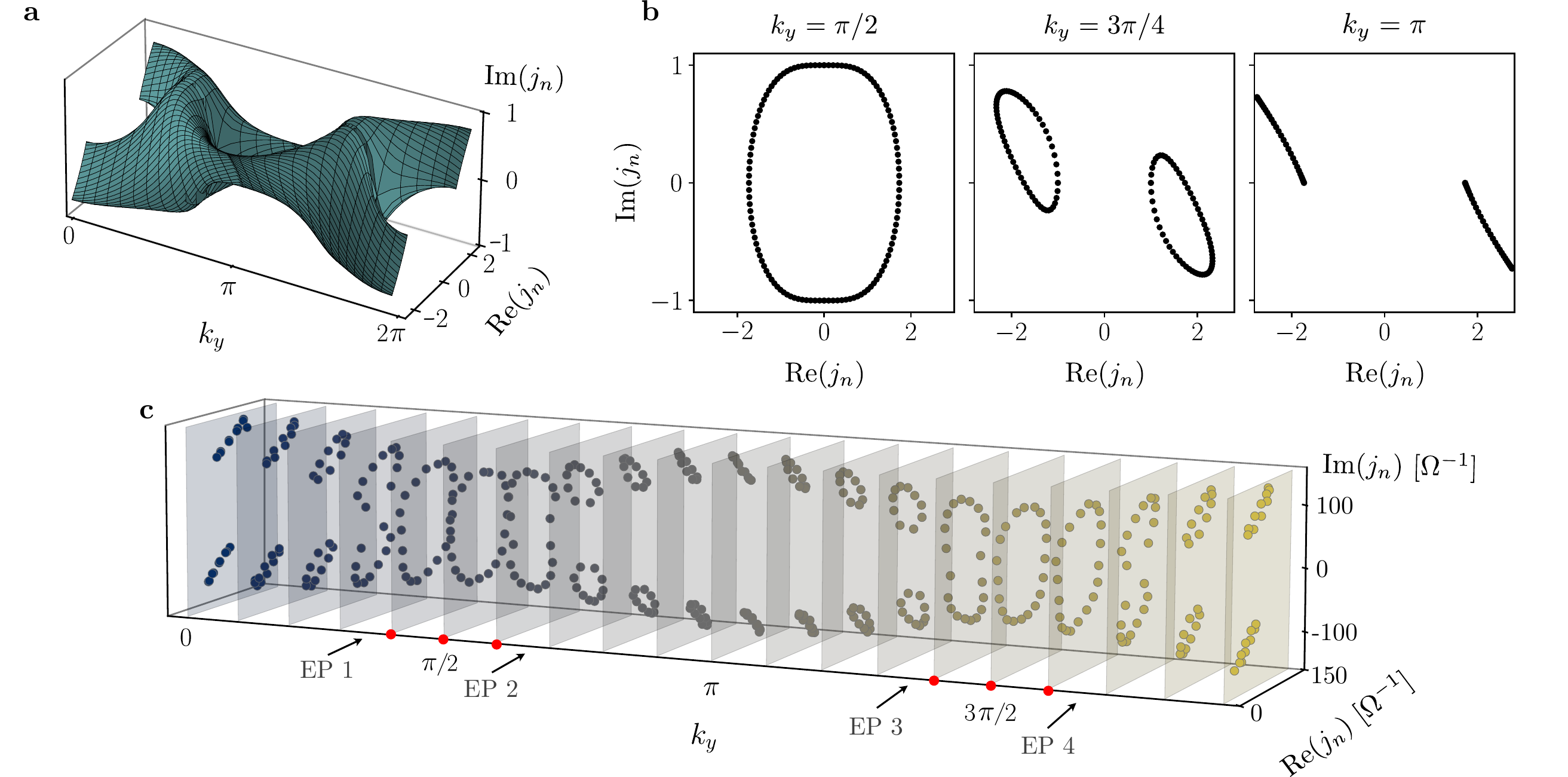}
\caption{
Representation of the circuit Laplacian spectra for periodic boundary conditions (PBC) in the complex plane as a function of $k_y$.
a) Theoretically computed spectrum showing a genus three surface with four exceptional points.
b) Slice plots obtained from a) for three different $k_y$.
c) Measured spectra, where each of the four transitions from a single circle in the complex plane to two circles marks an exceptional point (EP1--EP4) as a function of $k_y$. 
}
\label{fig: S2}
\end{center}
\end{figure*}

The circuit was fed by the sine wave generator of a lock-in amplifier (Zurich Instruments MFLI series). We worked with an input signal with an amplitude of $\text{900}\,\text{mV}$. The current was measured with an Oscilloscope (Rhode \& Schwarz RTM 1054) over a resistor (11 $\Omega$).
In order to measure the band structure of the circuit, we fed the current into a node via the resistor and we measured the voltage response due to the excitation at all nodes. For the voltage measurements, the same lock-in amplifier was used. We perform a Fourier transformation of the input current and voltage vector, for PBC in $x$ and $y$ direction and for OBC in $y$ direction only. Using Eq.~\eqref{eq:Greens Function Measurement}, we can reconstruct one column of the Green's function for all wave vectors. By repeating this procedure for all other \emph{inequivalent} nodes, the complete Green's function is accessible. For PBC only the different nodes in the unit cell are inequivalent, whereas for OBC all nodes in the non-periodic direction serve as current input for a set of measurements. After the Green's function is determined, we obtain the eigenvalues and eigenvectors by diagonalizing it. For details on the measurement process see Ref.~\onlinecite{PhysRevB.99.161114}.

\section{Representation of the Laplacian spectra in the complex plane}

We present an alternative representation of the complex spectra of the circuit Laplacian, both theoretical and experimental. For this, we consider $k_y$ as a parameter and plot for each fixed $k_y$ the set of eigenvalues in the complex plane $\mathrm{Re}(j_n)$-$\mathrm{Im}(j_n)$. This generates a two-dimensional closed surface, as shown in Fig.~\ref{fig: S2}~a) for PBC. As a function of $k_y$, the eigenvalues trace out either a single circle or two circles in the $\mathrm{Re}(j_n)$-$\mathrm{Im}(j_n)$ plane. The transitions between these two topologically distinct situations are exceptional points.
The planes $k_y=0$ and $k_y=\pi$ are special, because the path traced out in the complex plane as a function of $k_x$ is reciprocal on them. Therefore, it collapses into lines on which each non-end point is visited two times as $k_x$ is varied\cite{PhysRevB.99.201103}.

In summary, since we have two pairs of exceptional points, the two-dimensional eigenvalue surface in $k_y$-$\mathrm{Re}(j_n)$-$\mathrm{Im}(j_n)$ space has genus three\cite{lee2018tidal}.
We can plot the experimental spectra of $J(\omega_0)$ in a similar fashion, shown in Fig.~\ref{fig: S2}~c). We observe, for PBC, a clear from slices at constant $k_y$ with a single circle and two circles in the complex plane of eigenvalues. This strongly indicates the presence of exceptional points between these two slices. 

\section{Response of the reciprocal skin-effect circuit to bulk perturbations}
\label{sec: responsesupplementary}

\begin{figure*}[t]
\begin{center}
\includegraphics[width=\textwidth]{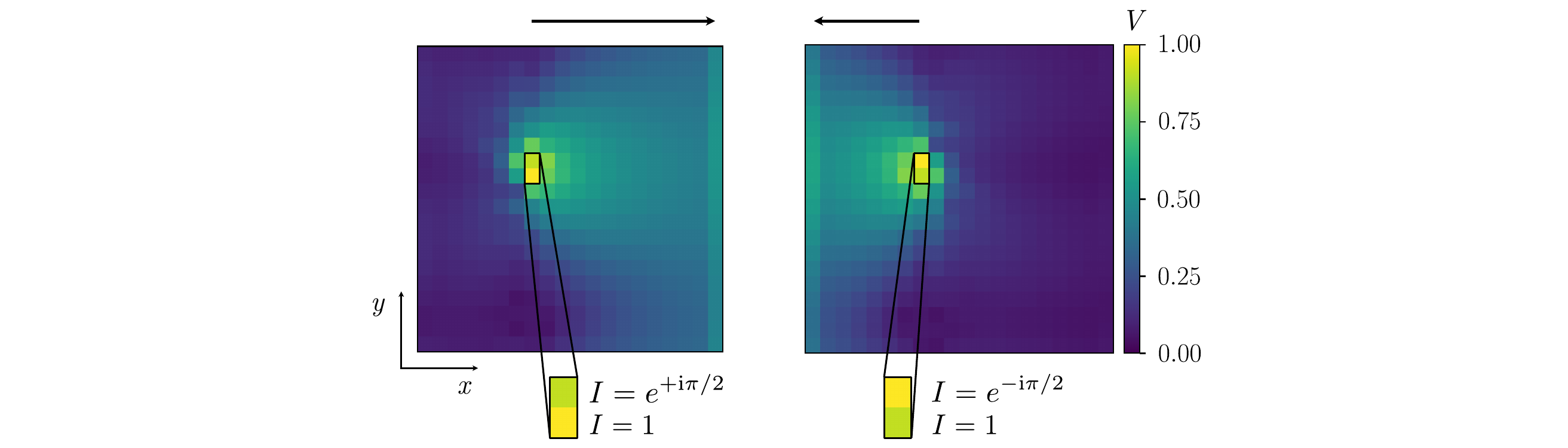}
\caption{
Reciprocal skin effect voltage response due to a localized bulk driving current with phase shift. The sites where the current is applied are framed in black, the zoom-ins show how the relative phase is implemented. Importantly, for a phase shift of $+\pi/2$ ($-\pi/2$) we find a nonlocal voltage response at the far right (left) side of the circuit with OBC, which is absent for phase shifts $0$ and $\pi$. The arrows show the direction of voltage accumulation.
}
\label{fig: S3}
\end{center}
\end{figure*}

A distinct feature of the reciprocal skin effect, which may lead to applications for polarization detection of electromagnetic waves, is the response to the circuit to perturbations in the bulk. From Eq.~\eqref{I2} we deduce that the voltage response at site $a$ to a driving current at frequency $\omega$ is given by
\begin{equation}
V_a=J^{-1}_{ab}(\omega)I_b.
\end{equation}
Since our circuit has finite resistivity, $J_{ab}(\omega)$ has no zero eigenvalues at real frequencies, and can therefore be readily inverted. To model the experimental situation, we consider PBC in $y$-direction and OBC in $x$-direction. We then apply a minimal driving current to two bulk sites aligned along the $y$-direction, where we assign the current at one site a $+\pi/2$ ($-\pi/2$) phase shift with respect to the current at the other site. We therefore expect the current to excite the circuit Laplacian eigenstates at $k_y = \pi/2$ ($k_y = 3\pi/2$), which are right (left) localized in $x$-direction due to the reciprocal skin effect. Figure~\ref{fig: S3} shows the theoretically calculated response of the circuit that is obtained by inverting the real-space version of the Hamiltonian given by Eq.~(2). We find that the two driving current patterns indeed lead to nonlocal voltage responses at the far right (left) of the circuit. The reciprocal skin effect could therefore be the basis of a polarization detection device for electromagnetic radiation.

%%%%%%%%%%%%%%%%%%%%%%%%%%%%%%%%%%%%%%%%%%%%%%%%%%%%

\cleardoublepage

%\bibliography{bibliography_new}


\end{document}